\documentclass[%
 reprint,
 amsmath,amssymb,
 aps,
]{revtex4-1}

\usepackage{graphicx}% Include figure files
\usepackage{mathtools}
\usepackage{dcolumn}% Align table columns on decimal point
\usepackage{bm}% bold math
\newenvironment{ABedit}{\begingroup\color{blue}}{\endgroup}
\newcommand{\bAB}{\begin{ABedit}}
\newcommand{\eAB}{\end{ABedit}}
\begin{document}

\preprint{APS/123-QED}

\title{Effect of reorientation statistics on torque response of self propelled particles}% Force line breaks with \\

\author{Benjamin Hancock}
 \email{bhancock@brandeis.edu}%Lines break automatically or can be forced with \\
\author{Aparna Baskaran}%
 \email{aparna@brandeis.edu}
\affiliation{%
 Martin Fisher School of Physics, Brandeis University, Waltham, MA 02453, USA
}%

\date{\today}% It is always \today, today,

\begin{abstract}
We consider the dynamics of self-propelled particles subject to external torques. Two models for the reorientation of self-propulsion are considered, run-and-tumble particles, and active Brownian particles. Using the standard tools of non-equilibrium statistical mechanics we show that the run and tumble particles have a more robust response to torques. This macroscopic signature of the underlying reorientation statistics can be used to differentiate between the two types of self propelled particles. Further this result might indicate that run and tumble motion is indeed the evolutionarily stable dynamics for bacteria.
\end{abstract}

\pacs{Valid PACS appear here}% PACS, the Physics and Astronomy
\maketitle

%\tableofcontents

\section{Introduction}
Self-propelled particles are inherently out of equilibrium objects that
consume energy at the scale of individual entities to produce persistent
self generated motion. Models of self propelled particles have become the
prototype theoretical system used to understand the physics of active
materials. These models are directly applicable to micron sized objects such
as motile bacteria and self-phoretic colloids \cite{Cates2012a, Saha2014,
Romanczuk2012, Marchetti2013, Palacci2010, Palacci2013,
Theurkauff2012,Golestanian2007, Buttinoni2013}.

There have been numerous theoretical investigations of the
collective behavior of active particles in the recent years. Phenomenology
uncovered includes cluster formation, phase separation, motility induced
segregation  \cite{Fily2012,McCandlish2012, Buttinoni2013, Redner2013,
Cates2014a, Schwarz2012, Arvind2012, Eli2014}, anomalous
behavior of mechanical properties such as pressure  \cite%
{Solon2014,Solon2014b} and rheology \cite{Fielding2011, Foffano2012}. Further, it has been established that the presence of external
forces \cite{Nash2010a, Tailleur2008a, Tailleur2009} or confinement \cite%
{Fily2014, Yang2014} dramatically alters the observed phenomenology in these
systems. One striking example has been rectification \cite{Galajda2007,
Reichhardt2013, reichardt2013a, Angelani2011,Tailleur2009, Wan2008}, in which asymmetric
barriers induce directed transport of self-propelled particles.

\indent Generally, self-propelled particles travel along a body axis $\hat{
\mathbf{u}}$ that has an intrinsic reorientation process.  Two model
classes of self-propelled particles that have been used extensively are
Run-and-Tumble particles (RTPs) and Active Brownian particles (ABPs). The
only difference in the dynamics of the two classes is the nature of their
reorientation process. The dynamics of an RTP consists of periods of
straight line motion followed by a sudden tumble event occurring at some
mean tumble rate $\alpha $. The tumble event completely decorrelates the
orientation and a new orientation is selected at random. There are several
examples of motile bacteria that follow run-and-tumble dynamics \cite{Eisenbach2005}, the
most famous example being \emph{Escherichia coli}.\cite{Berg2003}. The second class, ABPs, reorient their body axis through gradual rotational diffusion and this diffusion is Brownian with
rotational diffusion coefficient $D_{r}$. These dynamics have been observed
in experiments of synthetic self-phoretic colloids \cite{Palacci2010,
Palacci2013, Theurkauff2012,Golestanian2007, Buttinoni2013}.

For times much longer than the angular reorientation time, that is, $t\gg
\frac{1}{\alpha }$ for RTPs and $t\gg \frac{1}{D_{r}}$ for ABPs, both
classes of particles exhibit diffusive behavior. Equating the diffusivities
at this stage would yield identical phenomenology for both models. This idea
has been explored extensively in \cite{Cates2013, Solon2015, Tailleur2009, Cates2014a, Nash2010a, Cates2012a} where the authors find equivalence between the two models at
the level of a drift diffusion equation when $\alpha =(d-1)D_{r}$, where $d$
is the spatial dimension.

\begin{figure}[h!]
  \centering
  \includegraphics[width=8.8cm,height=3.5cm]{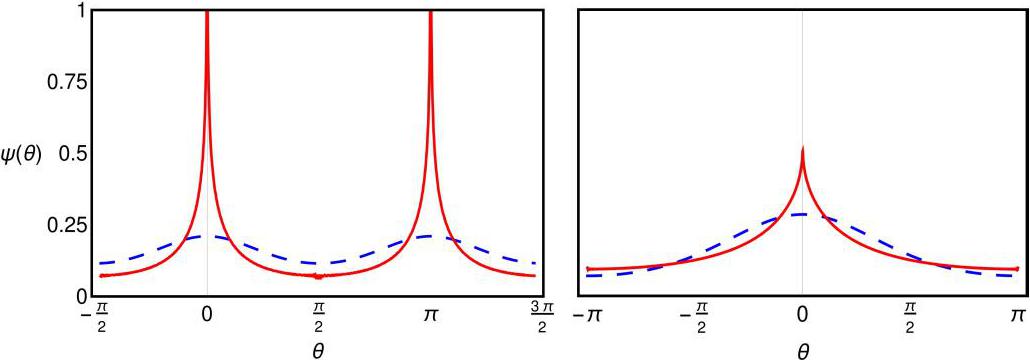}
\caption{(color online) Orientation probability distribution functions for RTPs (red/solid) and ABPs (blue/dashed) in the case of both a nematic aligning (left) field and a polar aligning field (right). RTP distributions are sharply peaked about the optimal direction while the ABP distribution is the smooth equilibrium result.}
\end{figure}

In this work we seek to identify signatures of the reorientation
statistics on the long time behavior of self-propelled particles. In
particular we ask the following question : If the swimming parameters are
equivalent, how does the long time behavior of ABPs and RTPs differ when
subject to external torques? External torques model gradient seeking
behavior exhibited by microorganisms such as chemotaxis \cite
{Eisenbach2005,Berg2003}, and aerotaxis \cite{Sokolov2009}. Such phenomena need not
be restricted to biological systems, as micron
sized platinum-gold rods have been shown to exhibit directed movement towards regions of higher
hydrogen-peroxide concentrations \cite{Hong2007}.  By using the framework of non-equilibrium
statistical mechanics we show that run-and-tumble particles align with
external torques (or gradients) much more robustly than active Brownian
particles, indicating that the reorientation behavior of bacteria is ideal for successful gradient seeking.

\section{Theory}

Let us consider non-interacting self propelled particles moving in
unconstrained 2D space. The statistical mechanics of this system is given in
terms of $\psi (\mathbf{r},\hat{\mathbf{u}},t)$, the probability of
finding a particle at a point $\mathbf{r}$ at a time $t$ oriented in a
direction $\mathbf{\hat{u}}$ $=\cos \theta \hat{x}+\sin \theta \hat{y}$ in
space. When the particle reorients its direction through rotational
diffusion, the dynamics of this probability distribution obeys a diffusion
equation of the form
\begin{equation} \label{1}
\begin{split}
\partial _{t}\psi (\mathbf{r},\hat{\mathbf{u}},t) =-\nabla \cdot
\lbrack v\hat{\mathbf{u}}&\psi (\mathbf{r},\hat{\mathbf{u}}%
,t)]+\nabla \cdot \lbrack D_{t}\nabla \psi (\mathbf{r},\hat{\mathbf{u%
}},t)]  \\
 &+\mathcal{\partial _{\theta }}[D_{r}\partial _{\theta }\psi (%
\mathbf{r},\hat{\mathbf{u}},t)]
\end{split}
\end{equation}
where $v$ is the self propulsion speed of the particle, $D_{t}$ is a
translational diffusion constant and $D_{r}$ is a rotational diffusion
constant. When the particle undergoes run and tumble dynamics, the
probability distribution obeys a master equation of the form%
\begin{equation}
\begin{split}
\partial _{t}\psi (\mathbf{r},\hat{\mathbf{u}},t)&=-\nabla \cdot
\lbrack v\hat{\mathbf{u}}\psi (\mathbf{r},\hat{\mathbf{u}}%
,t)]+\nabla \cdot \lbrack D_{t}\nabla \psi (\mathbf{r},\hat{\mathbf{u%
}},t)] \\
& -\alpha \psi (\mathbf{r},\hat{\mathbf{u}},t)+\frac{\alpha }{%
2\pi }\int d\hat{\mathbf{u}}^{\prime }\psi (\mathbf{r},\hat{%
\mathbf{u}}^{\prime },t)  \label{2}
\end{split}
\end{equation}%
where $\alpha $ is the tumble frequency. This description is valid when the
duration of a tumble is short compared to the time scales set by the self
propulsion and the tumble frequency of the particle. We are interested in
the dynamics of this system in the presence of uniform external torques. So,
in the following, we consider the homogeneous limit of Eqs. (\ref{1}-\ref{2})
in the presence of a $\theta $ dependent potential. The dynamical equations
of interest then are of the form

\begin{equation}
\partial _{t}\psi (\theta ,t)=\mathcal{\partial _{\theta }}[D_{r}\partial
_{\theta }\psi (\theta ,t)]+\frac{1}{\xi ^{r}}\partial _{\theta }[\partial
_{\theta }V(\theta )\psi (\theta ,t)]]  \label{3}
\end{equation}%
for active Brownian particles and
\begin{equation}
\begin{split}
\partial _{t}\psi (\theta ,t) =&-\alpha \psi (\theta ,t)+\frac{\alpha }{2\pi }
 \int d\theta ^{\prime }\psi (\theta ^{\prime },t)\\&+\frac{1}{\xi ^{r}}\partial
_{\theta }[\partial _{\theta }V(\theta )\psi (\theta ,t)]]  \label{4}
\end{split}
\end{equation}
for run and tumble particles. In the above $\xi ^{r}$ is a rotational
friction constant characteristic of the medium in which the particles move.

We will now consider different choices for the external potential and construct series solutions to Eqs. (\ref{3}-\ref{4}%
) of the form $\psi (\theta )=a_{0}+\sum_{n=1}^{\infty }a_{n}\cos (n\theta
)+b_{n}\sin (n\theta )$ for different choices of external potential $V\left(
\theta \right) $. Other useful measures to characterize the response of the system are the first two
moments of the orientation distribution, namely, the polarization $\mathbf{P}
$ defined as

\[
P_{\alpha}=\int d\theta \hat{u}_{\alpha}\psi (\theta )
\]%
and the nematic order parameter tensor $\overleftrightarrow{\mathbf{Q}}$

\[
Q_{\alpha \beta }=\int d\theta (\hat{u}_{\alpha }\hat{u}_{\beta }-\frac{1}{2}%
\delta _{\alpha \beta })\psi (\theta )
\]
These will be calculated as well.

\section{Nematic Torque}

First we consider an external potential of the form $V(\theta)=-\gamma \cos(2\theta)$. Such a field will induce a nematic aligning torque on the particles because of the two minima located at 0 and $\pi$. In this case a good characterization of the degree of order will be given by the component of the nematic order parameter tensor describing alignment in the $\hat{x}$ direction, namely $Q_{xx}$.

\begin{figure}[h!]
  \centering
  \includegraphics[width=8.4cm,height=5.8cm]{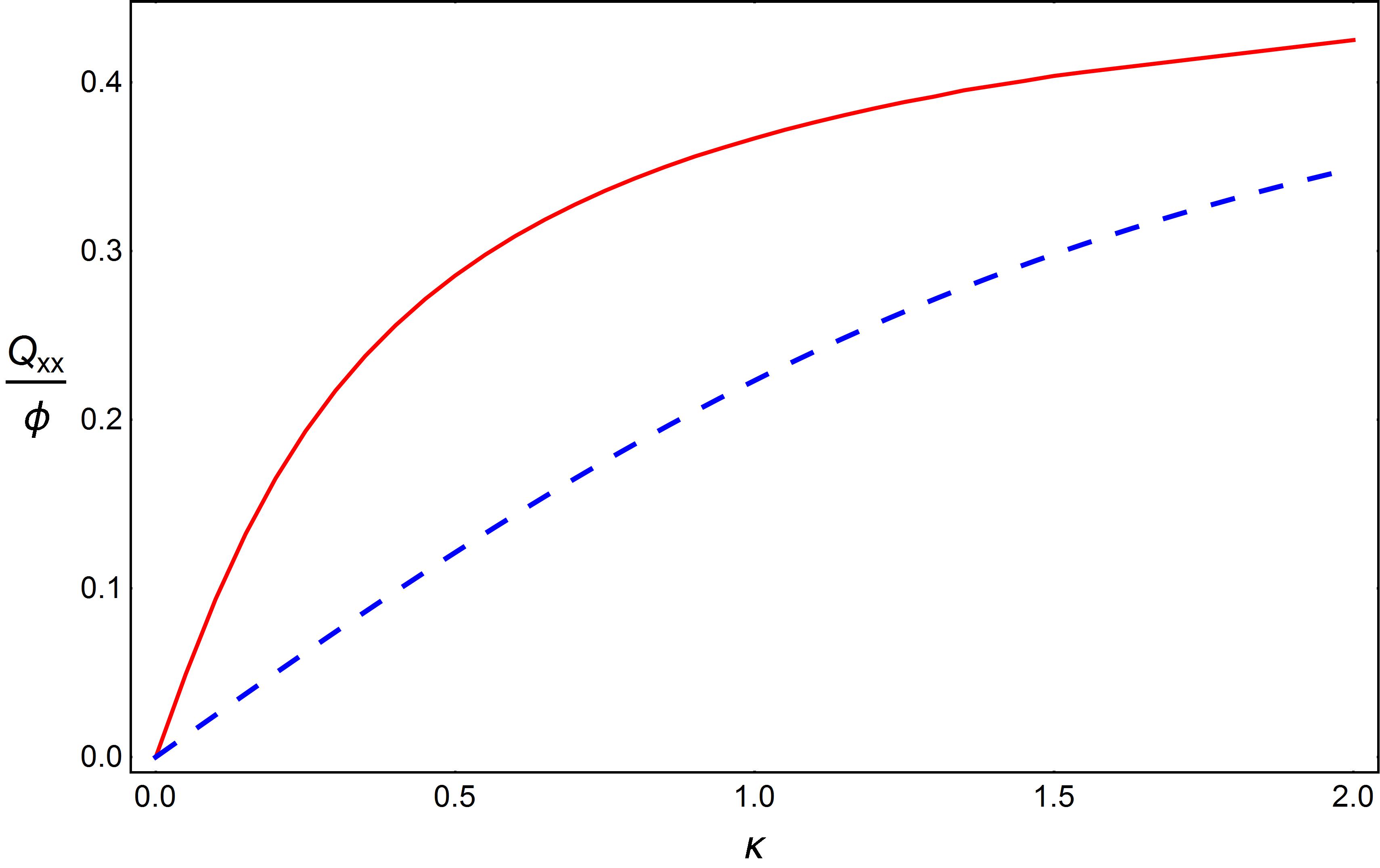}
\caption{(color online) RTPs exhibit higher nematic order than ABPs as shown by the nematic order parameter being higher for RTPs (red) than for ABPs (blue/dashed) }
\label{fig.2}
\end{figure}

\indent In the following, the parameter $\kappa$ is the ratio of the field strength of the external potential to the friction coefficient in units of the angular reorientation time and is given by  $\kappa=\frac{\gamma}{\xi^rD_r}$ for active Brownian particles and $\kappa=\frac{\gamma}{\xi^r\alpha}$ for run and tumble particles. Computing $Q_{xx}$ from (\ref{3}), one can show (see Appendix \ref{app:a} for details) that for active Brownian particles, we obtain
\begin{equation}
Q_{xx}=\frac{\phi}{2}\frac{I_1(\kappa)}{I_0(\kappa)} \label{5}
\end{equation}
which is the equilibrium result for a thermal nematogen at temperature $\xi^rD_r$ in the presence of an external field. Here the $I_i$'s are modified Bessel functions of the second kind \cite{nist}. For run and tumble particles, the exact value of the order parameter turns out to be a continued fraction of the form
\begin{equation}
Q_{xx}=\frac{\phi}{2}\cfrac{\kappa}{\frac{1}{2}+\cfrac{\kappa^2}{\frac{1}{4}+\cfrac{\kappa^2}{\frac{1}{6}+\cfrac{\kappa^2}{\frac{1}{8}+\dots}}}} \label{6}
\end{equation}
As seen in Fig.\ref{fig.2}, the nematic order parameter is larger for run-and-tumble particles.

We can also exactly compute the orientation distribution functions in the presence of this external torque. For active Brownian particles the distribution is of the form (Appendix \ref{app:a})
\begin{equation}
\psi(\theta)=\frac{e^{\kappa \cos(2\theta)}}{2 \pi I_0(\kappa)} \label{7}
\end{equation}
which is precisely the equilibrium result and is usually called a Von Mises distribution \cite{watson,evans}, the analog of the Gaussian distribution on a unit sphere. For run and tumble particles we obtain the following formally exact series solution
\begin{equation}
\psi (\theta )=a_{0}+\sum_{n=1}^{\infty }a_{2n}(\kappa)\cos (2n\theta
)\label{8}
\end{equation}
where the coefficients are given by
\[
a_2=\frac{2}{\pi}Q_{xx} \quad \textrm{and} \quad
     \begin{array}{lr}
       a_{2n}=a_{2n-4}-\frac{a_{2n-2}}{(2n-2)\kappa} & : n\geq2
       %a_4=2a_0-\frac{a_2}{2\kappa} &
     \end{array}
\]
The orientation distribution is much more peaked around the minima of the external potential for run and tumble particles than for active Brownian particles (see Fig.\ref{1}), another manifestation of the stronger response that RTPs exhibit to external torques. This more robust response to external torques is a consequence of the run-and-tumble search strategy being able to sample the optimal directions more efficiently than its Brownian counterpart, as the tumble directions are entirely random and unbiased by the initial orientation.

\section{Polar torque}
In this section we consider particles subject to an external potential of the form $V(\theta)=-\gamma \cos(\theta)$, i.e., a torque that tends to align their direction of motion along one direction in space. In this case, the relevant measure of the response of the system is the polarization. For active Brownian particles the polarization is found to be (Appendix \ref{app:a})
\begin{equation}
P_{x}=\phi \frac{I_1(\kappa)}{I_0(\kappa)} \label{9}
\end{equation}
while for run and tumble particles we have
\begin{equation}
P_{x}=\phi\cfrac{\kappa}{2+\cfrac{\kappa^2}{1+\cfrac{\kappa^2}{\frac{2}{3}+\cfrac{\kappa^2}{\frac{2}{4}+\dots}}}} \label{10}
\end{equation}

\noindent where $\kappa$ is as defined earlier. As in the nematic case in section 3, the result for the active Brownian particle is the same as that of equilibrium. The orientation distribution functions for the two classes are found to be
\begin{equation}
\psi(\theta)=\frac{e^{\kappa \cos(\theta)}}{2 \pi I_0(\kappa)} \label{11}
\end{equation}
for ABPs, while for RTPs
\begin{equation}
\psi (\theta )=a_{0}+\sum_{n=1}^{\infty }a_{n}(\kappa)\cos (n\theta
)\label{12}
\end{equation}
where the coefficients are given by
\[
a_1=\frac{1}{\pi}P_{x} \quad \textrm{and} \quad
     \begin{array}{lr}
       a_{n}=a_{n-2}-\frac{2a_{n-1}}{(n-1)\kappa} & : n\geq2
      % a_2=2a_0-\frac{2a_1}{\kappa} &
     \end{array}
\]
These results are plotted as a function of the dimensionless field strength $\kappa$ in Fig 3a and Fig 1.  Unlike in the nematic case, at the level of the moments of the distribution, the ABPs exhibit higher polar ordering than the RTPs even though the distribution itself is more sharply peaked about the optimal direction. A better measure of the response in this case is the percentage of particles that find the optimal direction. We find that in terms of this measure the response of the run-and-tumble particle is indeed much more robust (see Fig(\ref{3}b).

\begin{figure}[h!]
  \centering
  \includegraphics[width=9.2cm,height=4.0cm]{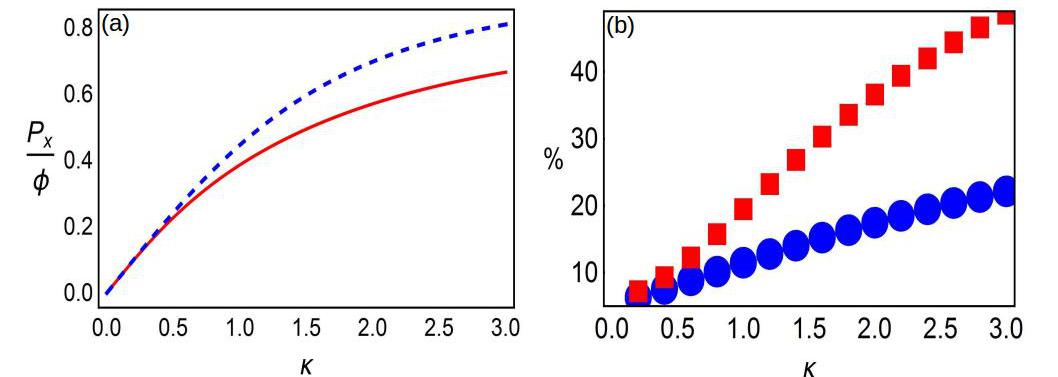}
\caption{(color online) (a): The polarization as  a function of the parameter $\kappa$. The polarization for RTPs (red/solid) is less than for ABPs (blue/dashed) (b):The percentage of particles oriented within some range $[\pi/18,\pi/18]$  as a function of the parameter $\kappa$. It is clear that the percentage of RTPs (red/square) oriented in that range is higher than ABPs (blue/circle).}
\label{fig.3}
\end{figure}

\section{Discussion}

\indent We have shown that the response of self-propelled particles to external torques is sensitive to the nature of the reorientation statistics of the propulsion direction. We find that run and tumble particles exhibit a more robust response to applied torques in the following sense. When there is one optimal direction for the particle's orientation, such as in the polar case, the fraction of particles that find the optimal direction is greater for those particles using a run-and-tumble search strategy, even though the average polarization is higher for particles undergoing Brownian rotational diffusion. For cases when there is more than one optimal direction for the particles to move in (such as the nematic case, where there exists 2 optimal directions) the run-and-tumble strategy always leads to better alignment as the particles are able to sample the different optimal directions better than through Brownian diffusion. These external torques model gradient seeking behavior exhibited by microorganisms and the more robust response of run-and-tumble particles might indicate a biological preference toward this search strategy.

Even though the torque response is dramatically different based on the reorientation statistics, a number of bulk phenomena such as phase separation and clustering are identical in the two classes of particles. This dichotomy can be understood by noting that the two models of self propulsion considered here have identical correlations but their response is very different. While the response of the Brownian particle has the conventional fluctuation-dissipation relationship to the correlation, the run-and-tumble particles do not have this property (see Appendix C). Therefore the nature of the reorientation statistics becomes important when considering external perturbations such as confining fores or aligning torques.

Finally we note that the exact expressions obtained in this work for run-and-tumble particles are given in terms of formal infinite series. These series solutions converge very slowly (see Appendix B) and any truncation at low orders drastically fails to capture the true nature of the solution. Therefore, low moment closure estimates, conventionally used in the literature of active particles, is unreliable for the case of run-and-tumble particles though they work very well for the case of active Brownian particles.

\section*{Acknowledgments}
 BH and AB acknowledge support from NSF-DMR-1149266, the Brandeis-MRSEC through NSF DMR-0820492 and NSF DMR-1420382. AB acknowledges KITP for hospitality through NSF-PHY11-25915 and fruitful conversations with Mike Cates and Julien Tailluer. BH also acknowledges support through IGERT DGE-1068620.

\appendix

\section{Derivation of order parameters and distribution functions}
\label{app:a}
In this appendix, the details of the derivation of steady state solutions to equations (\ref{3}-\ref{4}) for different external torques are given. Let us begin by considering the nematic torque. The statistical mechanics in this case is given by a Fokker-Planck equation for active Brownian particles
\begin{equation}
\partial _{t}\psi (\theta ,t)=\mathcal{\partial _{\theta }}[D_{r}\partial
_{\theta }\psi (\theta ,t)]+\frac{2\gamma}{\xi ^{r}}\partial _{\theta }[\sin(2\theta)\psi (\theta ,t)]] \label{a1}
\end{equation}
 and the following master equation for run and tumble particles 
\begin{equation}
\begin{split}
\partial _{t}\psi (\theta ,t)=&-\alpha \psi (\theta ,t)+\frac{\alpha }{2\pi }%
\int d\theta ^{\prime }\psi (\theta ^{\prime },t)\\ &+\frac{2\gamma}{\xi ^{r}}\partial
_{\theta }[\sin(2\theta)\psi (\theta ,t)]] \label{a2}
\end{split}
\end{equation}
As described earlier we search for series solutions $\psi (\theta )=a_{0}+\sum_{n=1}^{\infty }a_{n}\cos (n\theta
)+b_{n}\sin (n\theta )$. This Fourier decomposition is equivalent to the following moment expansion
\[
\psi(\theta)=\frac{1}{2\pi}(\phi+2 P_a \hat{u}_a+4 Q_{\alpha \beta}(\hat{u}_\alpha \hat{u}_\beta-\frac{1}{2}\delta_{\alpha \beta})+\dots)
\]
In this form it is apparent that $Q_{xx}=\frac{\pi}{2}a_2$. Using this with Eq.(\ref{a1}) we arrive at the following hierarchy of equations for active Brownian particles
\begin{equation}
\partial_{t}a_{0}=0
\end{equation}
\begin{equation}
\begin{array}{cc}
\partial_{t}a_{n}\delta_{n,m}=\frac{\gamma}{\xi^{r}}ma_{n}(\delta_{2,n+m}+sgn(m-n)\delta_{2,|m-n|})\\
-n^{2}D_{r}a_{n}\delta_{n,m}+\frac{2\gamma}{\xi^{r}}ma_{0}\delta_{2,m}
\end{array}
\end{equation}
\begin{equation}
\begin{array}{cc}
\partial_{t}b_{n}\delta_{n,m}=-\frac{\gamma}{\xi^{r}}mb_{n}(\delta_{2,n+m}+sgn(n-m)\delta_{2,|m-n|})\\-n^{2}D_{r}b_{n}\delta_{n,m}
\end{array}
\end{equation}
And similarly we find the corresponding hierarchy for run and tumble particles,
\begin{equation}
\partial_{t}a_{0}=0
\end{equation}
\begin{equation}
\begin{array}{cc}
\partial_{t}a_{n}\delta_{n,m}=\frac{\gamma}{\xi^{r}}ma_{n}(\delta_{2,n+m}+sgn(m-n)\delta_{2,|m-n|})\\
-\alpha a_{n}\delta_{n,m}+\frac{2\gamma}{\xi^{r}}ma_{0}\delta_{2,m}
\end{array}
\end{equation}
\begin{equation}
\begin{array}{cc}
\partial_{t}b_{n}\delta_{n,m}=-\frac{\gamma}{\xi^{r}}mb_{n}(\delta_{2,n+m}+sgn(n-m)\delta_{2,|m-n|})\\- \alpha b_{n}\delta_{n,m}
\end{array}
\end{equation}
First note that all odd coefficients are zero by the requirement the $\psi(\theta)=\psi(\theta+\pi)$. By truncating at arbitrary $a_{2n+2}$ or $b_{2n+2}$ and taking the steady state solution, one finds that all $a_{2n}$ or $b_{2n}$ couple to the zeroth coefficient $a_0$ or $b_0$. Since $b_0$ does not exist, the $b_{2n}$ vanish and by using an iterative procedure, we solve for $a_2$ and find for active Brownian particles
\begin{equation}
a_2=2 a_0\cfrac{\kappa}{2+\cfrac{\kappa^2}{4+\cfrac{\kappa^2}{6+\cfrac{\kappa^2}{8+\dots}}}} \label{a9}
\end{equation}
while for run and tumble particles we have
\begin{equation}
a_2=2a_0\cfrac{\kappa}{\frac{1}{2}+\cfrac{\kappa^2}{\frac{1}{4}+\cfrac{\kappa^2}{\frac{1}{6}+\cfrac{\kappa^2}{\frac{1}{8}+\dots}}}} \label{a10}
\end{equation}
Beyond Eq.(\ref{8}) no further analytic treatment for run and tumble particles is available. For active Brownian particles the continued fraction Eq.(\ref{a9}) is of the form of a Gauss continued fraction. In general we have the following identity for the ratio of modified Bessel functions
\begin{equation}
\frac{I_\nu(z)}{I_{\nu-1}(z)}=\cfrac{z}{2\nu+\cfrac{z^2}{2(\nu+1)+\cfrac{z^2}{2(\nu+2)+\cfrac{z^2}{2(\nu+3)+\dots}}}} \label{a11}
\end{equation}
One can find continued fraction representations for all the coefficients in our series solution and by using Eq.(\ref{a11}) one finds that
\begin{equation}
a_{2n}=\frac{I_{n}(\kappa)}{I_{0}(\kappa)}
\end{equation}
thus the orientation distribution function is given by the exact form
\begin{equation}
\psi(\theta)=\frac{1}{2\pi}+\frac{1}{\pi I_0(\kappa)}\sum_{n=1}^{\infty } I_n(\kappa)\cos(2 n \theta) \label{a13}
\end{equation}
With the help of the following identity, an example of a Jacobi-Anger expansion
\[
e^{z \cos(2\theta)}=I_0(z)+2\sum_{n=1}^{\infty } I_n(\kappa)\cos(2 n \theta)
\]
Eq.(\ref{a13}) is able to be summed exactly to the form of Eq.(\ref{7}). The same process holds for the polar aligning torque and repeating the calculation in that case yields Eq.(\ref{9})-(\ref{12})
\section{Truncation error and perturbation theory}
\begin{figure}[h!]
  \centering
  \includegraphics[width=8.2cm,height=5.8cm]{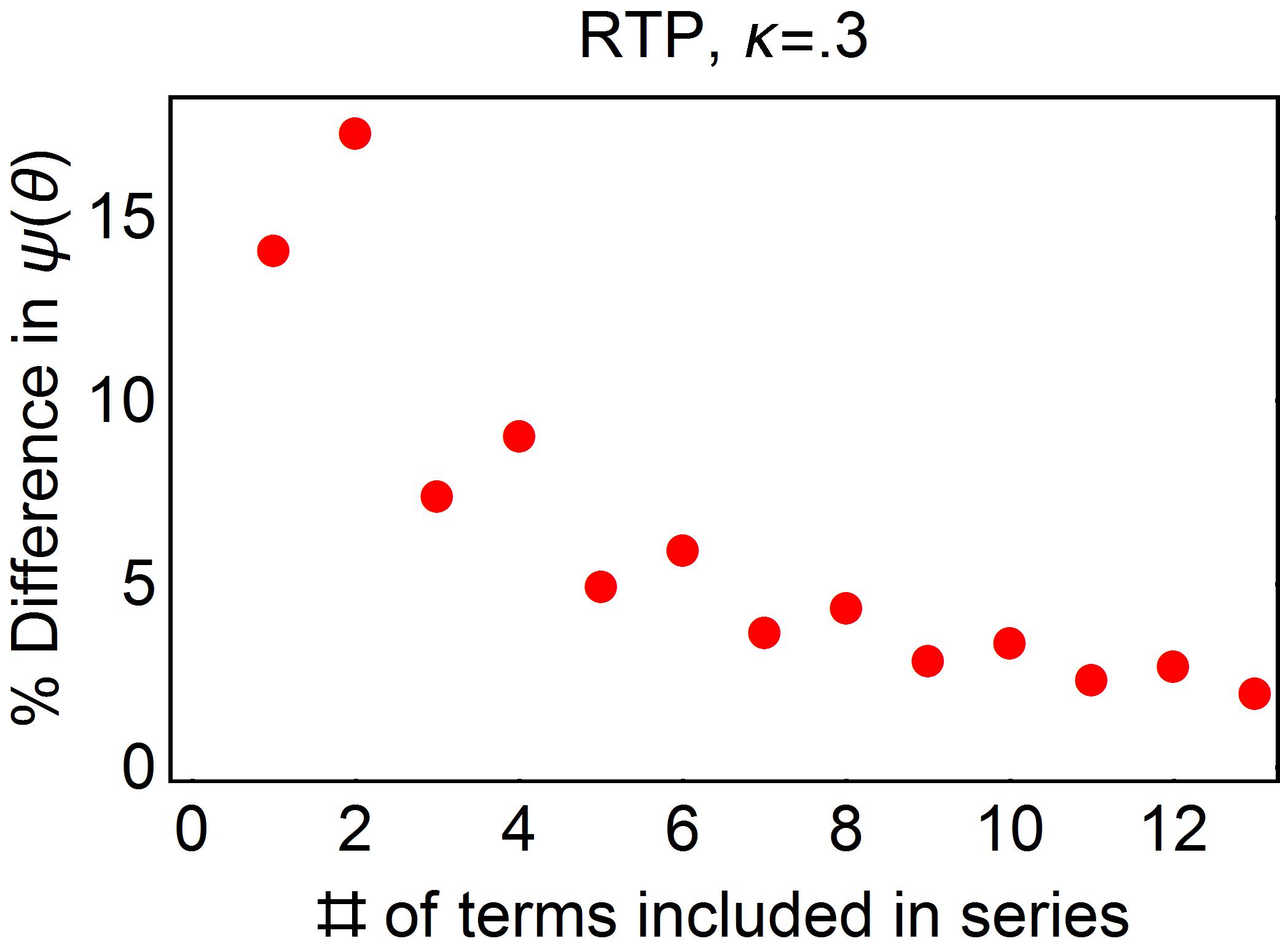}
\caption{(color online) For RTPs a large number of terms in the series must be included. Above shows the percent difference in the value of $\psi(\theta)$ whether truncating at the nth term or $n_{th}+1$ term. Here we examine the value at $\theta=0$. The first point on the horizontal axis represents the percent difference in $\psi(\theta)$ by truncating the series at $n=2$ and $n=3$ , the next point represents the percent difference when truncating the series at $n=3$ and $n=4$ this is continued up to the the last point which represents the percent difference by truncating the series at $n=14$ and $n=15$. }
\label{fig.4}
\end{figure}

As stated in the main text, the series solution for run and tumble particles is slowly converging and a low moment approximation is not valid in this case. One way to see this is to look at the $\%$ difference in the value of the function when you truncate the series solution at successive terms (see Fig.(\ref{fig.4})). Even in the weak field ($\kappa=.3$), a difference of about 15$\%$ is seen when truncating at the second or third terms in the series. For active Brownian particles, the same truncation yields a $1.4 \%$ difference, which shows that a low moment approximation is reasonable for ABPs. To obtain a $\%$ difference of this magnitude for RTPs, one must truncate the series at $n=300$. For both classes, the number of terms needed increases as $\kappa$ gets larger as should be expected.

\section{Relationship between Correlation and Response}
As another measure of the difference between the two classes of active particles one can compute the correlation and response functions. To illustrate, consider RTPs in the presence of a polar aligning torque.  To compute the correlation and response functions we will need the average of $\cos(\theta)$ , i.e., the polarization given by the first moment of the angular orientation distribution function.

\begin{equation}
\Bigl\langle\cos(\theta)\Bigr\rangle=\int d\theta\cos(\theta)[a_{0}+\sum_{n=1}^\infty a_{n}(\kappa)\cos(n\theta)]
\end{equation}
\[
=\pi a_{1}(\kappa)=2 \pi a_{0}\cfrac{\kappa}{2+\cfrac{\kappa^{2}}{1+\cfrac{\kappa^{2}}{\frac{2}{3}+\dots}}}=\cfrac{\kappa}{2+\cfrac{\kappa^{2}}{1+\cfrac{\kappa^{2}}{\frac{2}{3}+\dots}}}
\]
Where we used $a_{0}=1/2\pi$. The response function is giving by
\begin{equation}
R=\frac{\partial}{\partial\kappa}\Bigl\langle\cos(\theta)\Bigr\rangle=\frac{1}{2}-\frac{3}{4}\kappa^{2}+\mathcal{O}(\kappa^{3})
\end{equation}
The response function measures how the order parameter changes in response to a changing field. The correlation function is given by

\begin{equation}
C=\Bigl\langle\cos^{2}(\theta)\Bigr\rangle-\Bigl\langle\cos(\theta)\Bigr\rangle^{2}
\end{equation}
\[
=\frac{1}{2}+\frac{1}{2}a_{2}(\kappa)-a_{1}(\kappa)^{2};\,\,\,\,\,\,\,\, a_{2}=a_{1}\cfrac{\kappa}{1+\cfrac{\kappa^{2}}{\frac{2}{3}+\cfrac{\kappa^{2}}{\frac{2}{4}+\dots}}}
\]
\[
\implies C=\frac{1}{2}-\frac{\kappa^{4}}{4}+\mathcal{O}(\kappa^{5})
\]
So up to quadratic order in $\kappa$ we have
\begin{equation}
\frac{C}{R}=1+\frac{3}{2}\kappa^{2}+\dots
\end{equation}

\begin{figure}[h!]
  \centering
 \includegraphics[width=8.6cm,height=5.8cm]{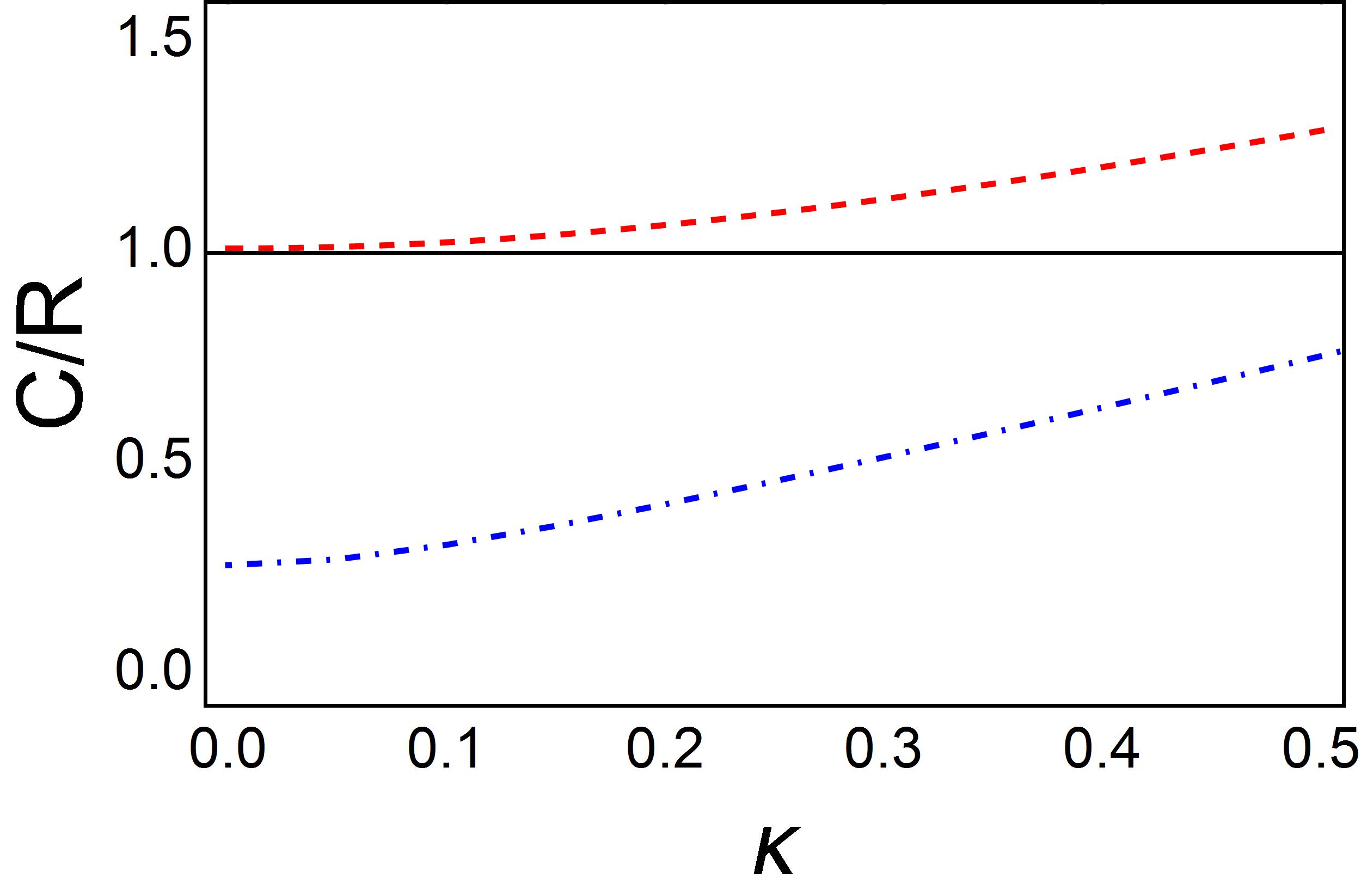}
\caption{(color online) The ratio of correlation to response for RTPs in the presence of a polar aligning torque (Red/Dashed) and a nematic aligning torque (Blue/Dot-Dashed) as a function of $\kappa$. The constant line at 1 represents the value of this ratio for ABPs}
\label{fig.5}
\end{figure}

Similary, for a nematic aligning field the relevant quantity is
\begin{equation}
\Bigl\langle\cos(2\theta)\Bigr\rangle=\int d\theta\cos(2\theta)[a_{0}+\sum_{n=1}^\infty a_{2n}(\kappa)\cos(2n\theta)]
\end{equation}
\[
=\pi a_{2}(\kappa)=2\pi a_{0}\cfrac{\kappa}{\frac{1}{2}+\cfrac{\kappa^{2}}{\frac{1}{4}+\cfrac{\kappa^{2}}{\frac{1}{6}+\dots}}}=\cfrac{\kappa}{\frac{1}{2}+\cfrac{\kappa^{2}}{\frac{1}{4}+\cfrac{\kappa^{2}}{\frac{1}{6}+\dots}}}
\]
Where we used $a_{0}=1/2\pi$. The response function is giving by
\begin{equation}
R=\frac{\partial}{\partial\kappa}\Bigl\langle\cos(2\theta)\Bigr\rangle=2-48\kappa^{2}+\mathcal{O}(\kappa^{3})
\end{equation}
The correlation function is given by

\begin{equation}
C=\Bigl\langle\cos^{2}(2\theta)\Bigr\rangle-\Bigl\langle\cos(2\theta)\Bigr\rangle^{2}
\end{equation}
\[
=\frac{1}{2}+\frac{1}{2}a_{4}(\kappa)-a_{2}(\kappa)^{2};\,\,\,\,\,\,\,\, a_{4}=a_{2}\cfrac{\kappa}{\frac{1}{4}+\cfrac{\kappa^{2}}{\frac{1}{6}+\cfrac{\kappa^{2}}{\frac{1}{8}+\dots}}}
\]
\[
\implies C=\frac{1}{2}-64 \kappa^{4}+\mathcal{O}(\kappa^{5})
\]
So up to quadratic order in $\kappa$ we have
\begin{equation}
\frac{C}{R}=\frac{1}{4}+6\kappa^2+\mathcal{O}(\kappa^{2})
\end{equation}

The exact expression for both aligning torques is explored numerically and shown in Fig.5. Repeating this process for ABPs would yield $C/R=1$ irrespective of the value of $\kappa$.

\bibliography{paperbib}

\end{document}